\documentclass[apj,twocolumn]{emulateapj}

\usepackage{amsmath}
\eqsecnum

\renewcommand\email\texttt

\def\spose#1{\hbox to 0pt{#1\hss}}
\def\lta{\mathrel{\spose{\lower 3pt\hbox{$\sim$}}
    \raise 2.0pt\hbox{$<$}}}
\def\gta{\mathrel{\spose{\lower 3pt\hbox{$\sim$}}
    \raise 2.0pt\hbox{$>$}}}

\begin{document} 

\slugcomment{\sc submitted to \it Astrophysical Journal Letters}
\shorttitle{\sc  Two New Milky Way Satellites} 
\shortauthors{}

\title{Big fish, small fish: Two New Ultra-Faint Satellites of the Milky Way}
\author{
V.\ Belokurov\altaffilmark{1},
M.\ G.\ Walker\altaffilmark{1}, 
N.\ W.\ Evans\altaffilmark{1}, 
G.\ Gilmore\altaffilmark{1},
M.\ J.\ Irwin\altaffilmark{1},
D.\ Just\altaffilmark{2},
S.\ Koposov\altaffilmark{1},
M.\ Mateo\altaffilmark{3},
E.\ Olszewski\altaffilmark{2},
L.\ Watkins\altaffilmark{1},
L.\ Wyrzykowski\altaffilmark{1}
}

\altaffiltext{1}{Institute of Astronomy, University of Cambridge,
Madingley Road, Cambridge CB3 0HA, UK;\email{vasily,walker,nwe@ast.cam.ac.uk}}
\altaffiltext{2}{Steward Observatory, University of Arizona, Tucson,
  AZ 85721, USA}
\altaffiltext{3}{Department of Astronomy, University of Michigan, 
Ann Arbor, MI 48109, USA}

\begin{abstract}
  We report the discovery of two new Milky Way satellites in the
  neighboring constellations of Pisces and Pegasus identified in data
  from the Sloan Digital Sky Survey. Pisces II, an ultra-faint dwarf
  galaxy lies at the distance of $\sim 180$ kpc, some 15$^\circ$ away
  from the recently detected Pisces I. Segue 3, an ultra-faint star
  cluster lies at the distance of 16 kpc. We use deep follow-up
  imaging obtained with the 4-m Mayall telescope at Kitt Peak National
  Observatory to derive their structural parameters. Pisces II has a
  half-light radius of $\sim$60 pc, while Segue 3 is twenty times
  smaller at only 3pc.
\end{abstract}


\section{Introduction}

Ultra-faint satellites of the Milky Way include dwarf galaxies
~\citep{Wi05,Zu06a,Zu06b,Be06a,Be07,Ir07,Be08} and star clusters
~\citep{Ko07}, as well as objects with intermediate properties
~\citep{Wal07,Be09, Ni09}. Defined by their extremely low surface
brightness, these systems could only be detected with a massive
multi-band imaging campaign like the Sloan Digital Sky Survey (SDSS).

In this {\it Letter}, we announce the discovery of two further Milky
Way satellites in the Southern Galactic portion of the SDSS SEGUE
survey. They lie in adjacent constellations and each extend only a
couple of arc-minutes on the sky. However, their heliocentric
distances differ by an order of magnitude, and, hence, so do their
physical sizes. We name the dwarf galaxy in the constellation of
Pisces, lying at the heliocentric distance of $\sim 180$ kpc and
measuring $\sim 120$ pc across, Pisces II. This is the second Galactic
stellar halo sub-structure in Pisces - the first, Pisces I was
announced earlier this year by ~\citet{Wat09}. Pisces I is much closer
and more dispersed on the sky: it is at least several degrees across
and lies at $\sim 80$ kpc. Our second discovery is a feeble cluster of
stars in the constellation of Pegasus. It has a half-light radius of 3
pc and lies at a heliocentric distance of 16 kpc. We name it Segue 3,
after SEGUE, the imaging survey in the data of which it was found.

In the analysis presented in this Letter we have extinction-corrected
all magnitudes using the maps of \citet{Sc98}.

\begin{figure*}[t]
\begin{center}
\includegraphics[width=0.9\textwidth]{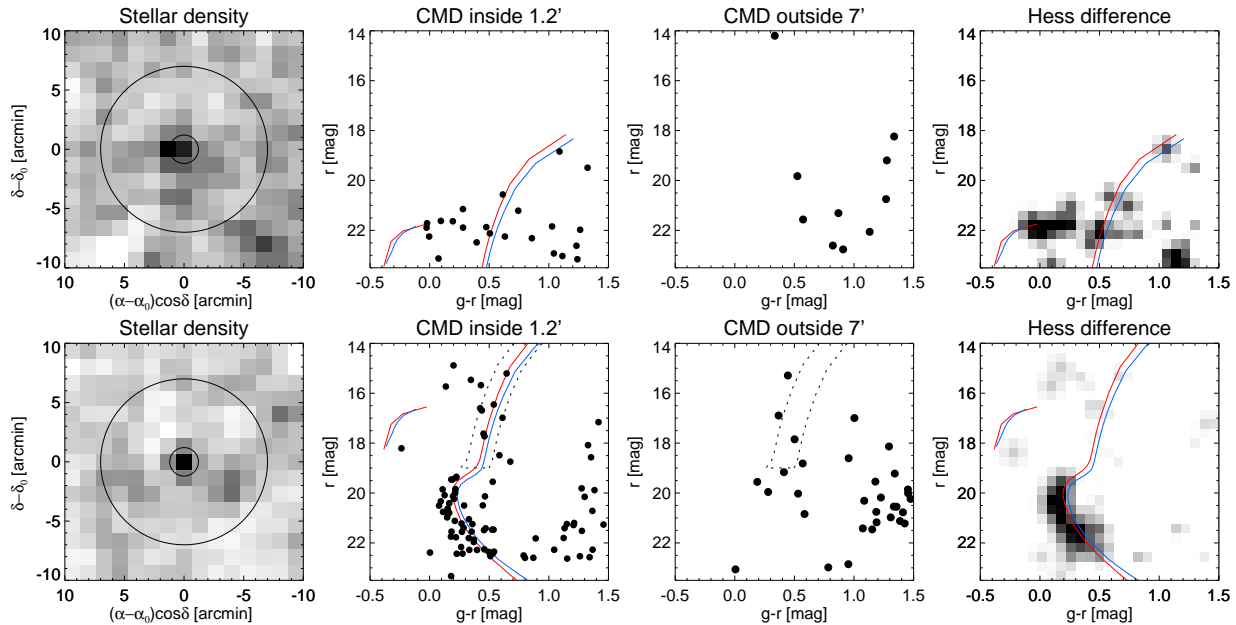}
\caption{The SDSS view of Pisces II (upper row) and Segue 3 (lower
  row): {\it Left:} Density of stars in the SDSS catalogue centered on
  the object. The stars in the $10^\prime \times 10^\prime$ area are
  binned into $15\times15$ bins and smoothed with a Gaussian with FWHM
  of 1.5 pixel. {\it Middle Left:} CMD of all stars in a circle of
  radius $1\farcm2$ (marked on the left panel), dominated by the
  satellite's members. {\it Middle Right:} Comparison CMD of stars
  within the annulus $7^\prime$ to $7\farcm1$ showing the
  foreground. {\it Right:} Difference in Hess Diagrams. Pisces
  II populations (top right) can be hinted at by the red giant branch
  and blue horizontal branch. Segue 3 (bottom right) shows an obvious
  main sequence. Ridge-lines of M92 (red, [Fe/H]$\sim$-2.3) and M13
  (blue, [Fe/H]$\sim$-1.55) are over-plotted. For Segue 3, we also
  show a mask built using M92 ridge-line to select possible red giant
  stars for luminosity calculation (lower middle left panel).}
\label{fig:fig1_sdss}
\end{center}
\end{figure*}
\begin{deluxetable}{lcc}
\tablecaption{Properties of Pisces II and Segue 3 \label{tbl:pars}}
\tablewidth{0pt} 
\tablehead{ \colhead{Parameter} & {Pisces II} & {Segue 3}} 
\startdata 
RA (J2000) &  $22:58:31 \pm 6$ & $21:21:31 \pm 4$ \\
Dec (J2000) & $+05:57:09 \pm 4$ & $+19:07:02 \pm 4$ \\
Galactic $\ell$ & $79.21^\circ$ & $69.4^\circ$ \\
Galactic $b$ & $-47.11^\circ$ & $-21.27^\circ$ \\
 $r_h$ (Plummer) & $1\farcm1 \pm 0\farcm1$ & $0\farcm65 \pm 0\farcm1$ \\
$\theta$ & $77^\circ \pm 12^\circ$ & $215^\circ \pm 20^\circ$\\
$e$ & $0.4 \pm 0.1$ & $0.3 \pm 0.2$\\
(m$-$M)$_0$ & $21\fm3$ & $16\fm1$ \\
 M$_{\rm tot,V}$ & $-5\fm0$ & $-1\fm2$
\enddata
\tablenotetext{*}{Magnitudes are accurate to $\sim \pm 0\fm5$ and are corrected for the Galactic foreground reddening.}
\label{tab:struct}
\end{deluxetable}

\section{Data and Discovery}

The SDSS imaging data are available through the latest Data Release 7
(DR7) in two parts: i) $\sim 8000$ square degrees of the main SDSS
field of view, mostly around the North Galactic Cap and ii) $\sim
3000$ square degrees of SEGUE imaging at low Galactic latitudes, with
large portions of the Southern Galactic sky
covered~\citep{Ya09}. These imaging data are produced in five
photometric bands, namely $u$, $g$, $r$, $i$, and $z$ and are
automatically processed through the same pipelines to measure
photometric and astrometric properties \citep{Ab09}. The two imaging
datasets differ in the continuity of the sky coverage and the amount
of Galactic reddening: while most of the SDSS has contiguous coverage
and is observed through minimal amounts of dust, SEGUE consists of
tens of long, $2.5^\circ$-wide stripes affected by various amounts of
Galactic extinction.

Both discontinuity in coverage and variable extinction complicate the
search for stellar over-densities by adding non-Poissonian noise to
the stellar density field. Nonetheless, applying our over-density
detection algorithm \citep{Be06a} to the SEGUE data immediately
yielded several promising candidates, which we are continuing to
follow up with deep imaging and spectroscopy. We have already
presented the first result, the discovery of the Segue 2 satellite
\citep{Be09}. Several more candidate objects were detected at similar
significance level. Pisces II was selected for follow-up, as it showed
tentative evidence for the presence of Blue Horizontal Branch (BHB)
stars. The case for Segue 3 was simpler: the object could actually be
seen on the SDSS images.

Fig.~\ref{fig:fig1_sdss} shows the view of Pisces II and Segue 3 as
seen by SDSS. The first of the four panels shows the density of all
objects classified as stars by the SDSS pipeline down to $r=23$. In
each case, there is a visible over-density at the center: Pisces II is
detected with significance \citep{Ko08} of $\sim 5$ and Segue 3 with
significance of $\sim 7$. The next two panels are the color-magnitude
diagrams (CMDs) of the object and the Galactic foreground around
it. It requires a lot of imagination to see the red giant branch or
indeed the blue horizontal branch of Pisces II. However, the Hess
difference in the right panel of Fig.~\ref{fig:fig1_sdss} shows that
there are over-densities of blue-ish and reddish stars that can be
interpreted as BHBs and RGBs at the heliocentric distance of $\sim
180$ kpc. The lower panel of Fig.~\ref{fig:fig1_sdss} presents the
SDSS data for Segue 3, which is clearly a simpler case: the main
sequence (MS) at $\sim 15$ kpc is obvious. It is, however, more
difficult to identify the RGB population in Segue 3 as there is no
obvious over-density in the Hess difference plot. In the absence of
spectroscopic data, we gauge the likely membership by placing a mask
that selects six potential members, at least one of which probably
belong to the Galactic foreground. There is also a lone blue star in
Segue 3, which looks slightly too faint to be classified as a BHB
unambiguously.

\begin{figure*}
\begin{center}
\includegraphics[width=0.9\textwidth]{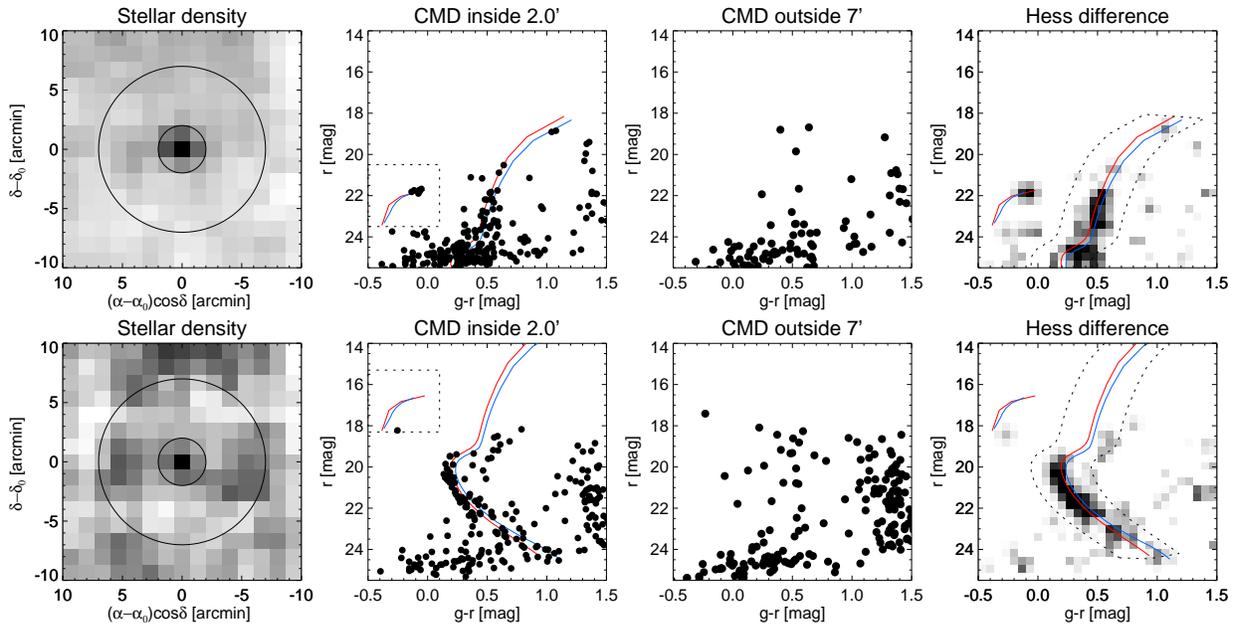}
\caption{The KPNO view of Pisces II (upper row) and Segue 3 (lower
  row). The panels are the same as in Figure 1, with the inner radius
  increased to $2^\prime$. The KPNO saturation limit is fainter than
  that of SDSS at $r\sim 18$, but it reaches $\sim 2$ magnitudes
  deeper. The CMD of Pisces II shows the most obvious improvement. In
  the KPNO data, strong RGB and clear BHB and blue straggler
  populations can now be seen. The dotted box in the middle left is
  used to pick out the likely BHB members. The dotted lines in the
  right panel outline the regions used to select the satellite
  members.}
\label{fig:fig1_kpno}
\end{center}
\end{figure*}
\begin{figure}
\begin{center}
\includegraphics[width=0.49\textwidth]{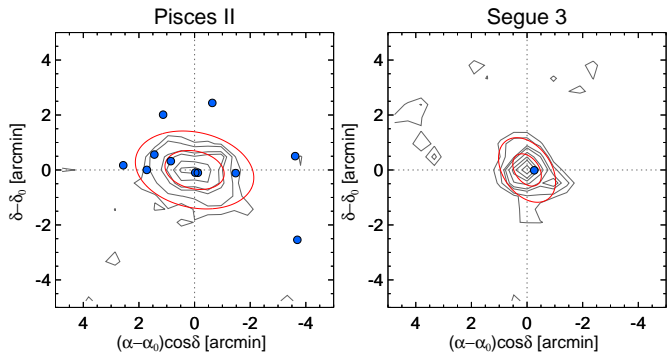}
\caption{Density contours (black) of candidate member stars selected
  from the KPNO data centered on each satellite. {\it Left:} Pisces II
  is mapped out with RGB stars. {\it Right:} Segue 3 is mapped out
  with MS stars. Contour levels are 3, 5, 8, 10, 15 $\sigma$ (also 20
  and 30 $\sigma$ for Segue 3) above the background. Red ellipses show
  Plummer isodensity contours corresponding to one and two
  half-radii. Blue dots mark the locations of the BHB candidate
  stars.}
  \label{fig:density}
\end{center}
\end{figure}

\begin{figure}
\begin{center}
\includegraphics[width=0.49\textwidth]{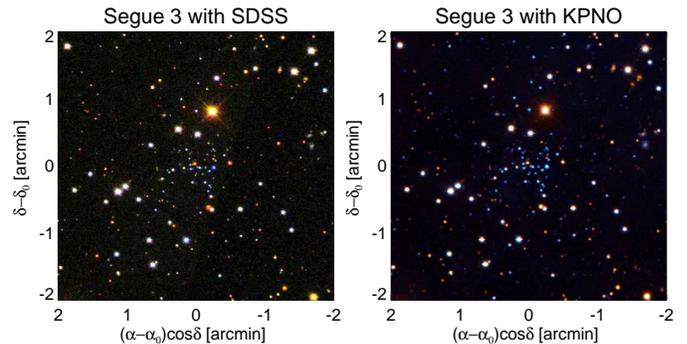}
\caption{Color images covering $4^\prime \times 4^\prime$ region
  centered on Segue 3 made with SDSS (left) and KPNO (right)
  data. SDSS image is made with $g,r $ and $i$ band frames. KPNO image
  is made with $g$ and $r$ band frames.}
\label{fig:seg_image}
\end{center}
\end{figure}

\section{Follow-Up Imaging and Structural Parameters}

During the nights of 19-24 September 2009, we obtained follow-up
photometry of Pisces II and Segue 3 using the MOSAIC camera at the 4-m
Mayall Telescope at Kitt Peak National Observatory, Arizona.  For both
objects, we observed a single $36\arcmin \times 36\arcmin$ field for
$3\times 900$ seconds in both $g$ and $r$ filters.  Each of MOSAIC's
eight detectors have $2048 \times 4096$ pixels with scale
$0.258\arcsec$ per pixel (unbinned).  For this run we enjoyed photometric
conditions and typical seeing of $\la 1\arcsec$.  We processed raw
MOSAIC frames using the set of IRAF-based procedures that have been
developed for the NOAO Deep Wide-Field Survey\footnote{Details of
these reduction procedures are available at
http://www.noao.edu/noao/noaodeep/ReductionOpt/frames.html}
\citep{Ja00}. These procedures include steps to correct for a pupil
ghost that is caused by reflections off the atmospheric dispersion
corrector.  Briefly, we used averaged dome flats to generate a
template image of the pupil ghost, which we then removed from the
master flat field image.  After dividing science frames by the master
flat field, we then scaled and subtracted the pupil template image
from science frames one by one.  Visual inspection confirms that these
steps successfully removed the pupil ghost from our science frames.
Data stacking and the production of catalogues was performed using a
general purpose pipeline for processing wide-field optical CCD
data~\citep{Ir01}. For each image frame, an object catalog was
generated and used to update the world coordinate system prior to
stacking each set of 3 frames.  A final set of object catalogs was
generated from the stacked images and objects were morphologically
classified as stellar or non-stellar (or noise-like).  The detected
objects in each passband were then merged by positional coincidence
(within $1^{\prime\prime}$) to form a combined $g,r$ catalogue and
photometrically calibrated on the SDSS system using stars in common.

{\it Pisces II} -- The KPNO photometry reaches at least 2 magnitudes
fainter than the SDSS and plays a crucial role in the identification
and analysis of Pisces II, as can be seen from
Fig.~\ref{fig:fig1_kpno}. The stellar over-density is enhanced, with a
significance of $\sim 8$. The CMD is now quite unambiguous, with both
RGB and BHB clearly visible together with a pile-up of stars around
the main sequence turn-off (MSTO). The very tight BHB can now be used
to estimate the distance modulus of the system (m$-$M)$_0 =
21.3$. Both CMD and the Hess difference panels of
Fig.~\ref{fig:fig1_kpno} show over-plotted the ridge-lines of the
globular clusters M92 and M13 from \citet{Cl05}. The RGB stars in
Pisces II seem to be described equally well by both and, hence, are
likely to possess metallicity in the range of $-2.3 < $[Fe/H]$ <
-1.55$ \citep{Ha96}. Note, however, that the two brightest RGB stars
are located right on top of the M13 ridgeline. To select all likely
members of Pisces II, we draw a wide mask shown in the right panel of
Fig.~\ref{fig:fig1_kpno}; we also select the BHB candidate stars with
a color-magnitude box shown in the middle left panel. Using these
selection cuts we can now measure the structural parameters of Pisces
II and its luminosity. The density contours of the RGB and MSTO stars
are shown in the left panel of Fig.~\ref{fig:density}. To measure the
satellite's half-light radius, ellipticity and position angle, we
follow the procedure outlined in \citet{Ma08}. From the locations of
candidate MSTO and RGB stars we measure a half-light radius of
$1\farcm1$ (or $\sim$ 60 pc at a distance of $\sim$ 180 kpc) with
noticeable ellipticity. The results of this maximum likelihood fit are
reported in Table~\ref{tab:struct}.

To estimate the total luminosity of Pisces II, we i) integrate the
flux inside the mask shown in the right-most panel of
Figure~\ref{fig:fig1_kpno} within $3^{\prime}$ to get V$=16.6$ mag;
ii) subtract an estimate of background contamination of $17.4$ mag;
iii) add the contribution from the BHBs and possible blue stragglers
(BS) of V $=19.4$ to get $M_{\rm V}=-4.2$. To account for the missing
flux in the fainter stars, we use the luminosity function of the Ursa
Minor dSph, which we integrate within $3.7 < $V$ < 13$ to get 0.8
mag. This can be compared to 0.5 mag correction if the luminosity
function of a typical globular cluster is used (see
e.g. \citet{Ni10}). Out of all ingredients
contributing to the total luminosity, the exact number of RGB members
of Pisces II carries the largest uncertainty. Our final estimate of
the luminosity of Pisces II is $M_{\rm V}=-5\pm 0.5$ mag.

{\it Segue 3} -- This satellite is one of the very few recent
discoveries that can be seen seen directly in the SDSS
images. Figure~\ref{fig:seg_image} shows color images of
4$^{\prime}\times4^{\prime}$ area centered on Segue 3 made with $g,r$
and $i$ SDSS frames and $g$ and $r$ KPNO frames. SDSS and KPNO data
are of comparable quality, but the KPNO frames are integrated longer
and hence the faint objects are detected and measured with greater
accuracy. On both images, a central concentration of bluish stars
belonging to Segue 3 can be seen, albeit with a somewhat irregular
distribution. The KPNO photometry presented in the lower panel of
Figure~\ref{fig:fig1_kpno} reveals a rather tight main sequence, with
a clear turn-off, which we use to estimate the distance modulus of
(m$-$M)$_0 = 16.1$. Of the two ridgelines over-plotted on the
satellite's CMD, M92, with [Fe/H]$\sim$-2.3, clearly provides a better
match. We use the KPNO data to derive the structural parameters of
Segue 3 by following the procedure mentioned above. The half-light
radius of Segue 3 is $0 \farcm 65$ and, overall, its distribution of
stars is circular (see Table~\ref{tab:struct}), although there are
some irregularities in the density profiles as can be seen from the
right panel of Figure~\ref{fig:density} most likely due to the small
numbers of stars. We calculate the luminosity of Segue 3 in two steps,
using both SDSS and KPNO data. For the MS members fainter than $r=19$,
we integrate the flux within $2^{\prime}$ inside the mask shown in the
bottom right panel of Figure~\ref{fig:fig1_kpno} and subtract the
estimate of the background contamination (outside $2^{\prime}$) to get
V$=16.6$. For the RGB members ($r<19$), we integrate flux inside a
narrower mask shown in the middle-left panel of
Figure~\ref{fig:fig1_sdss}, which, after the background subtraction,
gives V$=15.15$. So, the total V$=14.9$, or $M_{\rm V}=-1.2$. This is
accurate to not less than 0.5 mag and can be better constrained when
the true RGB members are identified through spectroscopic follow-up.

\section{Discussion and Conclusions}

Pisces~II is close on the sky to Pisces~I, discovered as an
overdensity of RR Lyraes by \citet{Wat09} and confirmed
spectroscopically by \citet{Ko09}. But, at a heliocentric distance of
$\sim 180$ kpc, Pisces~II is almost twice as far as away.  The extent
of Pisces~I is probably considerable, at least as judged from the RR
Lyrae populations (see Figure 12 of Watkins et al. 2009). This might
imply the break-up of a substantial satellite galaxy moving on a
radial orbit, in which case it is natural to interpret Pisces~II as a
further fragment or companion.  Pisces~II is very similar in
morphology, size and luminosity to a number of recent discoveries such
as Hercules, Leo~IV and Leo~V~\citep{Be07, Be08}. They also all lie at
similar heliocentric distances of $\sim 150$ kpc. All four have
extended BHB populations enshrouding them.

Segue~3 is a very close relative of Koposov~1 and 2, the ultrafaint
star clusters at distances of $\sim 50$ kpc~\citep{Ko07}. All three
objects have a similar size ($\sim 3$ pc), luminosity $M_{\rm V}\sim
-1$ and contain only a few tens of stars. Unlike Koposov~1 and 2,
Segue~3 is much closer, and might even be a part of the
Hercules-Aquila Cloud~\citep{Be07b,Wat09}.  The evolution of such
objects is known to proceed with pronounced mass segregation.  The
very few heavy stars sink to the centre, and the lighter stars are
ejected to form a diffuse corona. It would be interesting to verify
this prediction with spectrocopic surveys of the objects.

\acknowledgments 

Funding for the SDSS and SDSS-II has been provided by the Alfred P.
Sloan Foundation, the Participating Institutions, the National Science
Foundation, the U.S. Department of Energy, the National Aeronautics
and Space Administration, the Japanese Monbukagakusho, the Max Planck
Society, and the Higher Education Funding Council for England. The
SDSS Web Site is http://www.sdss.org/.  VB thanks the Royal Society
for financial support. SK, MAW, LLW and LW all thank the Science and
Technology Facilities Council of the UK for funding. EO acknowledges
NSF grant AST-0807498; MM acknowledges NSF grant AST-0808043


\end{document}